\documentstyle[aps,preprint,eqsecnum]{revtex}

\begin{document}

\title{Propagating torsion from first principles}
\author {Alberto Saa}

\address{Departamento de Matem\'atica Aplicada\\
       IMECC -- UNICAMP, C.P. 6065, \\
13081-970 Campinas, SP, Brazil}

\maketitle

\begin{abstract}
A propagating torsion model is derived from the requirement of
compatibility between minimal action principle and minimal coupling
procedure in Riemann-Cartan spacetimes. In the proposed model,
the trace of the torsion tensor is derived from a scalar potential
that determines the volume element of the spacetime. The equations
of the model are write down for the vacuum and for various types 
of matter fields. Some of their properties  are discussed. In particular,
we show that gauge fields can interact minimally with the torsion
without the braking of gauge symmetry.
\end{abstract}

\newpage

\section{Introduction}

Many ``connection-dynamic'' theories of gravity with propagating torsion
have been proposed in the last decades. Contrary to the usual
Einstein-Cartan (EC) gravity\cite{hehl}, in such theories one could 
in principle
have long-range torsion mediated interactions.
In the same period, we have also witnessed a spectacular progress in the
experimental description of the solar system\cite{will}. 
Many important tests using
the parameterized post-Newtonian (PPN) formalism have been performed. Tight
limits for the PPN parameters have been establishing and 
several alternatives
theories to General Relativity (GR)
 have been ruled out. Indeed, such solar system experiments
and also observations of the binary pulsar $1913+16$ offer strong evidence
that the metric tensor must not deviate too far from the predictions of
GR\cite{will}. 
Unfortunately, the situation with respect to the torsion tensor is
much more obscure. The interest in experimental consequences of propagating
torsion models has been revived recently\cite{CF,hamm}.
Carroll and Field\cite{CF} have examined the
observational consequences of propagating torsion in a wide class of
models involving scalar fields. They conclude that for reasonable models
the torsion must decay quickly outside matter distribution, leading to
no long-range interaction which could be detected experimentally.
Nevertheless, as also stressed by them, this does not mean that torsion
has not relevance in Gravitational Physics.
 
Typically, in
propagating torsion models the Einstein-Hilbert action is modified in 
order to induce a differential equation for the torsion tensor, allowing for
non-vanishing torsion configurations to the vacuum. In almost all cases
a dynamical scalar  field is involved, usually related to the torsion
trace or pseudo-trace. Such modifications are introduced in a
rather arbitrary way; terms are added to the Lagrangian in
order to produce previously desired differential 
equations for the torsion tensor.

The goal of this paper is to present a propagating torsion model
obtained from first principles of EC theory. By exploring some basic
features of the Einstein-Hilbert action in spacetimes with torsion  we
get a model with a new and a rather intriguing type of propagating torsion
involving a non-minimally coupled scalar field.
We write and discuss the metric and torsion equations for the vacuum 
and in the presence of different matter fields. 
 Our model does not belong to
the large class of models studied in \cite{CF}.
The work is organized as follows. Section II is a brief revision of
Riemann-Cartan (RC)
 geometry, with special emphasis to the concept of parallel volume
element. In the Section III, we show how a propagating torsion model
arises from elementary considerations on the compatibility between
minimal action principle and minimal coupling procedure. The Section
IV is devoted to study of the proposed model in the vacuum and in presence of
various type of matter. Section V is left to some concluding
remarks.

\section{RC manifolds and parallel volume elements}

A RC spacetime is a differentiable four dimensional
manifold endowed with a metric tensor $g_{\alpha\beta}(x)$ and with a 
metric-compatible connection $\Gamma_{\alpha\beta}^\mu$, which is 
non-symmetrical in its lower indices. We adopt in this work 
${\rm sign}(g_{\mu\nu})=(+,-,-,-)$.
The anti-symmetric part of the
connection defines  a new tensor, the torsion tensor,
\begin{equation}
S_{\alpha\beta}^{\ \ \gamma} = \frac{1}{2}
\left(\Gamma_{\alpha\beta}^\gamma-\Gamma_{\beta\alpha}^\gamma \right).
\label{torsion}
\end{equation}
The metric-compatible connection can be written as 
\begin{equation}
\Gamma_{\alpha\beta}^\gamma = \left\{_{\alpha\beta}^\gamma \right\} 
- K_{\alpha\beta}^{\ \ \gamma},
\label{connection}
\end{equation}
where $\left\{_{\alpha\beta}^\gamma \right\}$ are
 the usual Christoffel symbols 
and $K_{\alpha\beta}^{\ \ \gamma}$ is the
contorsion tensor, which is given in terms of the torsion tensor by 
\begin{equation}
K_{\alpha\beta}^{\ \ \gamma} = - S_{\alpha\beta}^{\ \ \gamma} 
+ S_{\beta\ \alpha}^{\ \gamma\ } - S_{\ \alpha\beta}^{\gamma\ \ }.
\label{contorsion}
\end{equation}
The connection (\ref{connection}) is used to define the covariant derivative of
vectors,
\begin{equation}
 D_\nu A^\mu = \partial_\nu A^\mu + \Gamma_{\nu\rho}^\mu A^\rho,
\label{covariant}
\end{equation}
and 
it is also important to our purposes to introduce the covariant derivative of
a density $f(x)$,
\begin{equation}
D_\mu f(x) = \partial_\mu f(x) - \Gamma^\rho_{\rho\mu}f(x).
\end{equation}

The contorsion tensor (\ref{contorsion}) can be covariantly 
split in a traceless part and in a trace,
\begin{equation}
K_{\alpha\beta\gamma} = \tilde{K}_{\alpha\beta\gamma} - 
\frac{2}{3}\left( 
g_{\alpha\gamma} S_\beta  - g_{\alpha\beta} S_\gamma
\right),
\label{decomposit}
\end{equation}
where $\tilde{K}_{\alpha\beta\gamma}$  is the traceless part and $S_\beta$ is 
the trace of the torsion tensor, $S_\beta = S^{\ \ \alpha}_{\alpha\beta}$.
In four dimensions the traceless part $\tilde{K}_{\alpha\beta\gamma}$
can be also decomposed in a pseudo-trace and a part with vanishing 
pseudo-trace, but for our purposes (\ref{decomposit}) is sufficient.
The curvature tensor is given by:
\begin{equation}
\label{curva}
R_{\alpha\nu\mu}^{\ \ \ \ \beta} = \partial_\alpha \Gamma_{\nu\mu}^\beta
- \partial_\nu \Gamma_{\alpha\mu}^\beta 
+ \Gamma_{\alpha\rho}^\beta \Gamma_{\nu\mu}^\rho 
- \Gamma_{\nu\rho}^\beta \Gamma_{\alpha\mu}^\rho .
\end{equation}
After some algebraic manipulations we get the following expression for the 
scalar of curvature $R$, obtained from suitable contractions of (\ref{curva}),
\begin{equation}
R\left(g_{\mu\nu},\Gamma^\gamma_{\alpha\beta}\right) = 
g^{\mu\nu} R_{\alpha\mu\nu}^{\ \ \ \ \alpha} =
{\cal R} - 4D_\mu S^\mu + \frac{16}{3}S_\mu S^\mu - 
\tilde{K}_{\nu\rho\alpha} \tilde{K}^{\alpha\nu\rho},
\label{scurv}
\end{equation}
where ${\cal R}\left(g_{\mu\nu},\left\{_{\alpha\beta}^\gamma \right\}\right)$ 
is the Riemannian scalar of curvature, calculated from the
Christoffel symbols.

In order to define a general covariant volume element in a manifold, it is
necessary to introduce a density quantity $f(x)$ which will compensate
the Jacobian that arises from the transformation law 
of the usual volume element
$d^4x$ under a coordinate transformation,
\begin{equation}
d^4x \rightarrow f(x) d^4x = d{\rm vol}.
\end{equation}
Usually, the density $f(x) = \sqrt{-g}$ is took to this purpose. However,
there are
natural properties that a volume element shall exhibit.
In a Riemannian manifold, the usual covariant volume element
\begin{equation}
d{\rm vol} = \sqrt{-g}\, d^4x,
\label{vele}
\end{equation}
is parallel, in the sense that the scalar density $\sqrt{-g}$ obeys
\begin{equation}
{\cal D}_\mu\sqrt{-g} = 0,
\end{equation}
where ${\cal D}_\mu$ is the covariant derivative defined using the
Christoffel symbols. 
One can infer that the volume element (\ref{vele}) is not parallel
when the spacetime is not torsionless, since
\begin{equation}
D_\mu\sqrt{-g}= \partial_\mu\sqrt{-g} - \Gamma^\rho_{\rho\mu}\sqrt{-g} = 
-2 S_\mu\sqrt{-g},
\end{equation}
as it can be checked using Christoffel symbols properties. This is the
main point that we wish to stress, it will be the basic argument to our claim
that the usual volume element (\ref{vele}) is not the most appropriate one 
in the presence of torsion, as it will be discussed in the next section.

The question
that arises now is if it is possible to define a 
parallel volume element in RC manifolds. 
In order to do it,
one needs to find a  density $f(x)$ such that 
$D_\mu f(x)=0$. Such density exists only if the trace
of the torsion tensor, $S_\mu$, can be obtained
from a scalar potential\cite{saa1}
\begin{equation}
S_\beta(x) = \partial_\beta \Theta(x),
\label{pot}
\end{equation}
and in this case we have $f(x)=e^{2\Theta}\sqrt{-g}$, and
\begin{equation}
d{\rm vol} = e^{2\Theta}\sqrt{-g} \,d^4x,
\label{u4volume}
\end{equation}
that is the parallel RC volume element,
or in another words, the volume element (\ref{u4volume}) is compatible
with the connection in RC manifolds  obeying (\ref{pot}). 
It is not usual to find in the literature applications where volume
elements different from the canonical one are used. 
Non-standard
 volume elements have been used
in the characterization of half-flats solutions of
 Einstein equations\cite{volu},
in the description of field theory on Riemann-Cartan 
spacetimes\cite{saa1,saa2} and of dilatonic gravity\cite{saa4},
and in the study of some aspects of BRST symmetry\cite{AD}.
In our case the
new volume element appears naturally; in the same way that we require
compatibility conditions between the metric tensor and the linear
connection we can do it for the connection and volume element.

With
the volume element (\ref{u4volume}), we have the following generalized 
Gauss' formula
\begin{equation}
\int d{\rm vol}\, D_\mu V^\mu = 
\int d^4x \partial_\mu e^{2\Theta}\sqrt{-g} V^\mu =\ 
{\rm surface\ term},
\label{gauss}
\end{equation}
where we used 
that 
\begin{equation}
\label{gammacontr}
\Gamma^\rho_{\rho\mu}=\partial_\mu\ln e^{2\Theta}\sqrt{-g}
\end{equation}
under the hypothesis (\ref{pot}). It is easy to see that one cannot have a
generalized Gauss' formula of the type (\ref{gauss}) if the torsion does not
obey (\ref{pot}). We will return to discuss the actual role of the condition
(\ref{pot}) in the last section.

\section{Minimal coupling procedure and minimal action principle}

As it was already said, our model arises from elementary considerations
on the minimal coupling procedure and minimal action principle.
Minimal coupling procedure (MCP) provides us with an useful rule to get
the equations for any physical field on non-Minkowskian manifolds starting
from their versions of Special Relativity (SR). When studying classical
fields on a non-Minkowskian manifold $\cal X$ we usually require that the
equations of motion for such fields have an appropriate SR limit. There are,
of course, infinitely many covariant equations on $\cal X$ with the same
SR limit, and MCP solves this arbitrariness by saying that the relevant
equations should be the ``simplest'' ones. MCP can be heuristically formulated
as follows. Considering the equations of motion for a classical field in
the SR, one can get their version for a non-Minkowskian spacetime $\cal X$
by changing the partial derivatives by the $\cal X$ covariant ones and the
Minkowski metric tensor by the $\cal X$ one. MCP is also 
used for the classical and quantum
analysis of gauge fields, where the gauge field is to be interpreted
as a connection, and it is in spectacular agreement with
experience for QED an QCD.

Suppose now that the SR equations of motion for a classical field follow
from an action functional via minimal action principle (MAP). It is natural to
expect that the equations obtained by using MCP to the SR equations
coincide with the Euler-Lagrange equations of the action obtained 
via MCP of the SR one. This can be better visualized with the help of the
following diagram\cite{saa5}
\setlength{\unitlength}{1mm}
$$
\addtocounter{equation}{1}
\newlabel{diagr}{{3.1}{3.1}}
\hspace{106pt}
\begin{picture}(52,28)
\put(3,20) {$ {\cal C}_{ {\cal L}_{\rm SR} }$}
\put(7,18){\vector(0,-1){9}}
\put(3,5){$ E({\cal L}_{\rm SR}) $}
\put(45,20){${ \cal C_{L_X} }$ }
\put(40,5){$ E({\cal L}_{\cal X})$}
\put(47,18){\vector(0,-1){9}}
\put(12,22){\vector(1,0){30}}
\put(17,7){\vector(1,0){22}}
\put(24,24){${\scriptstyle \rm MCP}$}
\put(27,9){${\scriptstyle \rm MCP}$}
 \put(8,13){${\scriptstyle \rm MAP}$}
\put(48,13){${\scriptstyle \rm MAP}$}
\end{picture}
\hspace{116pt}\raise 7ex \hbox{(\theequation)}
$$
where $E({\cal L})$ stands to the Euler-Lagrange equations for the
Lagrangian $\cal L$, and ${\cal C}_{\cal L}$ is the equivalence class 
of Lagrangians, ${\cal L}'$ being equivalent to $\cal L$ if 
$E({\cal L}')=E({\cal L})$. 
We restrict ourselves to the case of non-singular Lagrangians.
The diagram (\ref{diagr}) is verified in GR. We say that MCP is
compatible with MAP if (\ref{diagr}) holds. We stress that if (\ref{diagr})
does not hold we have another arbitrariness to solve, one needs to choose 
one between two equations, as we will shown with a simple example.

It is not difficulty to check that in general MCP is not compatible with MAP
when spacetime is assumed to be non-Riemannian. 
Let us examine for simplicity
the case of a massless scalar field $\varphi$ in the frame of Einstein-Cartan
gravity\cite{saa1}. The equation for $\varphi$ in SR is
\begin{equation}
\partial_\mu\partial^\mu\varphi=0,
\label{e2}
\end{equation}
which follows from the extremals of the action
\begin{equation}
\label{act}
S_{\rm SR} = 
\int d{\rm vol}\, \eta^{\mu\nu}\partial_\mu\varphi\partial_\nu\varphi.
\end{equation}
Using MCP to (\ref{act}) one gets
\begin{equation}
\label{act1}
S_{\cal X} = \int d{\rm vol}\, g^{\mu\nu}
\partial_\mu\varphi\partial_\nu\varphi,
\end{equation}
and using the Riemannian volume element for $\cal X$, $
d{\rm vol} = \sqrt{g}d^nx$, we get the following equation from the
extremals of (\ref{act1})
\begin{equation}
\label{aa22}
\frac{1}{\sqrt{g}}\partial_\mu \sqrt{g}\partial^\mu\varphi = 0.
\end{equation}
It is clear that (\ref{aa22}) does not coincide in general with the
equation obtained via MCP of (\ref{e2})
\begin{equation}
\label{e3}
\partial_\mu\partial^\mu\varphi + \Gamma^\mu_{\mu\alpha}
\partial^\alpha\varphi =
\frac{1}{\sqrt{g}}\partial_\mu \sqrt{g}\partial^\mu\varphi
+ 2 \Gamma^\mu_{[\mu\alpha]} \partial^\alpha\varphi = 0.
\end{equation}
We have here an ambiguity, the equations (\ref{aa22}) and (\ref{e3}) are in
principle equally acceptable ones, to choose one of them corresponds to choose
as more fundamental the equations of motion or the action formulation from
MCP point of view. As it was already said, we do not have such ambiguity
when spacetime is assumed to be
a Riemannian manifold. This
is not a feature of massless scalar fields, all matter fields have the
same behaviour in the frame of Einstein-Cartan gravity. 

An accurate analysis of the diagram (\ref{diagr}) reveals that the source
of the problems of compatibility between MCP and MAP is the volume element
of $\cal X$. The necessary and sufficient condition to the validity of
(\ref{diagr}) is that the equivalence class of Lagrangians  
${\cal C}_{\cal L}$ be preserved under MCP. With our definition of
equivalence we have that
\begin{equation}
\label{class}
{\cal C}_{ {\cal L}_{\rm SR} } \equiv \left\{ {\cal L}'_{\rm SR}|
 {\cal L}'_{\rm SR} -  {\cal L}_{\rm SR} = \partial_\mu V^\mu \right\},
\end{equation}
where $V^{\mu}$ is a vector field. The application of MCP to the
divergence $\partial_\mu V^\mu$ in (\ref{class}) gives $D_\mu V^\mu$,
and in order to the set
\begin{equation}
\left\{ {\cal L}'_{\cal X}|
 {\cal L}'_{\cal X} -  {\cal L}_{\cal X} = D_\mu V^\mu \right\}
\end{equation}
be an equivalence class one needs to have a Gauss-like law like
(\ref{gauss}) associated to
the divergence $D_\mu V^\mu$. 
As it was already said in Section II, the necessary and sufficient
condition to have such a Gauss law is that the trace of the torsion
tensor obeys (\ref{pot}).

With the use of the parallel volume element in the
action formulation for EC gravity we can have qualitatively
different predictions. The scalar of curvature (\ref{scurv})
involves terms quadratic in the torsion.
Due to (\ref{pot}) such quadratic terms will provide a differential
equation for $\Theta$, what will allow for non-vanishing torsion
solutions for the vacuum. 
As to the matter fields, the
use of the parallel volume element, besides of guarantee
that the diagram (\ref{diagr}) holds, brings also qualitative changes.
For example, it is possible to have a minimal interaction between
Maxwell fields and torsion preserving gauge symmetry. The next section
is devoted to the study of EC equations obtained by using the 
parallel volume element (\ref{u4volume}).

\section{The model}

Now, EC gravity will be reconstructed by
using the results of the previous sections. Spacetime will be 
 assumed to be a Riemann-Cartan
manifold with the parallel volume element (\ref{u4volume}), and of course,
it is implicit the restriction that the trace of the torsion tensor is
derived from a scalar potential, condition (\ref{pot}). 
With this hypothesis, EC theory of gravity will predict new effects, and they
will be pointed out in the following subsections.

\subsection{Vacuum equations}

According to our hypothesis,
in order to get the EC gravity equations we will assume that they 
can be obtained from an Einstein-Hilbert  action using the scalar of
curvature (\ref{scurv}), the condition (\ref{pot}), and the
volume element (\ref{u4volume}),
\begin{eqnarray}
\label{vaction}
S_{\rm grav} &=& -\int d^4x e^{2\Theta} \sqrt{-g} \, R   \\
&=&-\int d^4x e^{2\Theta} \sqrt{-g} \left( 
{\cal R} + \frac{16}{3} \partial_\mu\Theta \partial^\mu \Theta 
- \tilde{K}_{\nu\rho\alpha} \tilde{K}^{\alpha\nu\rho}
\right) + {\rm surf. \ terms}, \nonumber
\end{eqnarray}
where the generalized Gauss' formula (\ref{gauss}) was used.

The equations for the $g^{\mu\nu}$, $\Theta$, and 
$\tilde{K}_{\nu\rho\alpha}$ fields follow from the extremals of the action
(\ref{vaction}).
The variations of $g^{\mu\nu}$ and $S_{\mu\nu}^{\ \ \rho}$ are assumed to
vanish in the boundary.
The equation $\frac{\delta S_{\rm grav}}{\delta\tilde{K}_{\nu\rho\alpha}} =0$
implies that $\tilde{K}^{\nu\rho\alpha} =  0$,
$\frac{\delta S_{\rm grav}}{\delta\tilde{K}_{\nu\rho\alpha}}$ standing for the
Euler-Lagrange equations for
${\delta\tilde{K}_{\nu\rho\alpha}}$.
 For the other equations we have
\begin{eqnarray}
\label{1st}
-\frac{e^{-2\Theta}}{\sqrt{-g}} 
\left.\frac{\delta }{\delta g^{\mu\nu}}S_{\rm grav}
\right|_{\tilde{K}=0} &=& {\cal R}_{\mu\nu}
-2D_\mu \partial_\nu\Theta \nonumber \\
&&-\frac{1}{2}g_{\mu\nu}
\left( 
{\cal R} + \frac{8}{3}\partial_\rho\Theta \partial^\rho \Theta 
-4 \Box \Theta
\right) = 0,   \\
-\frac{e^{-2\Theta}}{2\sqrt{-g}} 
\left.\frac{\delta }{\delta \Theta}S_{\rm grav}
\right|_{\tilde{K}=0} &=& 
{\cal R} + \frac{16}{3}\left( 
\partial_\mu\Theta \partial^\mu \Theta -
\Box \Theta \right) =0, \nonumber
\end{eqnarray}
where 
${\cal R}_{\mu\nu}
\left(g_{\mu\nu},\left\{_{\alpha\beta}^\gamma \right\}\right)$ 
is the usual Ricci tensor, calculated using the
Christoffel symbols, and  $\Box = D_\mu D^\mu$.

Taking the trace of the first equation of (\ref{1st}),
\begin{equation}
{\cal R} + \frac{16}{3}\partial_\mu\Theta \partial^\mu \Theta = 
6\Box\Theta,
\end{equation}
and using it,  one finally obtains
the equations for the vacuum, 
\begin{eqnarray}
\label{vacum0}
{\cal R}_{\mu\nu} &=& 2D_\mu\partial_\nu \Theta 
- \frac{4}{3} g_{\mu\nu}\partial_\rho\Theta \partial^\rho \Theta
= 2D_\mu S_\nu - \frac{4}{3}g_{\mu\nu}S_\rho S^\rho, \nonumber \\
\Box \Theta &=& \frac{e^{-2\Theta}}{\sqrt{-g}}
\partial_\mu e^{2\Theta}\sqrt{-g}\partial^\mu\Theta = D_\mu S^\mu = 0, \\ 
\tilde{K}_{\alpha\beta\gamma} &=& 0. \nonumber
\end{eqnarray}

The vacuum equations (\ref{vacum0})  
point out new features of our model. It is
clear that torsion, described by the last two equations,
 propagates. 
The torsion mediated interactions are not of
contact type anymore. The traceless tensor $\tilde{K}_{\alpha\beta\gamma}$
is zero for the vacuum, and only the trace $S_\mu$ can be non-vanishing
outside matter distributions. As it is expected, the gravity field
configuration for the vacuum is determined only 
by boundary conditions, and if 
due to such conditions we have that $S_\mu=0$, our equations reduce to the
usual vacuum equations, $S_{\alpha\gamma\beta}=0$, and 
${\cal R}_{\alpha\beta}=0$. Note that this is the case if one considers 
particle-like solutions (solutions that go to zero asymptotically). 
Equations (\ref{vacum0}) are valid only to the exterior region of the
sources. For a discussion to the case with sources see \cite{H1}.

The first term in the right-handed side of the first equation 
of (\ref{vacum0}) appears
to be non-symmetrical under the change $(\mu\leftrightarrow\nu)$,
but in fact it is symmetrical as one can see using (\ref{pot}) and
the last equation of (\ref{vacum0}). Of course that if 
$\tilde{K}_{\alpha\beta\gamma}\neq 0$ such term will be non-symmetrical,
and this is the case when fermionic fields are present, as we will see.

It is not difficult to generate solutions for (\ref{vacum0})
starting from the well-known solutions of the minimally coupled
scalar-tensor gravity\cite{saa6}.

\subsection{Scalar fields}

The first step to introduce matter fields in our discussion 
will be the description of
scalar fields on RC manifolds.
In order to do it, we will use MCP according to Section II.
For a massless scalar field one gets
\begin{eqnarray}
\label{scala}
S &=& S_{\rm grav} + S_{\rm scal} = -\int  \,d^4xe^{2\Theta}\sqrt{-g}
\left(R -\frac{g^{\mu\nu}}{2} \partial_\mu\varphi \partial_\nu \varphi 
\right)\\
&=&-\int d^4x e^{2\Theta} \sqrt{-g} \left( 
{\cal R} + \frac{16}{3} \partial_\mu\Theta \partial^\mu \Theta 
- \tilde{K}_{\nu\rho\alpha} \tilde{K}^{\alpha\nu\rho}
-\frac{g^{\mu\nu}}{2} \partial_\mu\varphi \partial_\nu \varphi
\right), \nonumber 
\end{eqnarray}
where surface terms were discarded.
The equations for this case are obtained by varying (\ref{scala}) with
respect to $\varphi$, $g^{\mu\nu}$, $\Theta$, and 
$\tilde{K}_{\alpha\beta\gamma}$. As in the vacuum case, the equation 
$\frac{\delta S}{\delta \tilde{K}}=0$
implies $\tilde{K}=0$. Taking it into
account we have
\begin{eqnarray}
\label{e1}
-\frac{e^{-2\Theta}}{\sqrt{-g}} \left. 
\frac{\delta S}{\delta\varphi}
\right|_{\tilde{K}=0} &=&  \frac{e^{-2\Theta}}{\sqrt{-g}}\partial_\mu
e^{2\Theta}\sqrt{-g}\partial^\mu\varphi 
=\Box \varphi = 0, \nonumber \\
-\frac{e^{-2\Theta}}{\sqrt{-g}} \left. 
\frac{\delta S}{\delta g^{\mu\nu}}
\right|_{\tilde{K}=0} &=& {\cal R}_{\mu\nu} 
- 2 D_\mu S_\nu - \frac{1}{2} g_{\mu\nu}
\left( 
{\cal R} + \frac{8}{3}S_\rho S^\rho - 4 D_\rho S^\rho
\right) \nonumber \\ 
&&-\frac{1}{2} \partial_\mu \varphi \partial_\nu\varphi 
+ \frac{1}{4} g_{\mu\nu}\partial_\rho \varphi \partial^\rho \varphi = 0, \\
-\frac{e^{-2\Theta}}{2\sqrt{-g}} \left. 
\frac{\delta S}{\delta \Theta}
\right|_{\tilde{K}=0} &=& {\cal R} + 
\frac{16}{3}\left( S_\mu S^\mu - D_\mu S^\mu\right) 
 -\frac{1}{2} \partial_\mu\varphi \partial^\mu\varphi = 0. \nonumber
\end{eqnarray}
Taking the trace of the second equation of (\ref{e1}),
\begin{equation}
{\cal R} + \frac{16}{3} S_\mu S^\mu = 6 D_\mu S^\mu +
\frac{1}{2} \partial_\mu\varphi \partial^\mu \varphi,
\end{equation}
and using it, we get the following 
set of equations for the massless scalar case
\begin{eqnarray}
\label{aa}
\Box \varphi &=&  0, \nonumber \\
{\cal R}_{\mu\nu} &=& 2D_\mu S_\nu - \frac{4}{3}g_{\mu\nu} S_\rho S^\rho 
+\frac{1}{2} \partial_\mu\varphi \partial_\nu\varphi, \\
D_\mu S^\mu &=& 0, \nonumber \\
\tilde{K}_{\alpha\beta\gamma} &=& 0. \nonumber
\end{eqnarray}

As one can see, the torsion equations have the same form than the ones
 of the vacuum case (\ref{vacum0}). Any
contribution to the torsion will be due to boundary conditions, and not due
to the scalar field itself. 
It means that if such boundary conditions imply that $S_\mu=0$, the
equations for the fields $\varphi$ and $g_{\mu\nu}$ will be the same ones
of the GR.
One can interpret this by saying that, 
even feeling the torsion (see the second equation of (\ref{aa})),
massless scalar fields do not produce it. Such  behavior is 
compatible with the idea that torsion must be governed by spin distributions.

However, considering massive scalar fields,
\begin{eqnarray}
S_{\rm scal} = \int  \,d^4xe^{2\Theta}\sqrt{-g}
\left(\frac{g^{\mu\nu}}{2} \partial_\mu\varphi \partial_\nu \varphi 
-\frac{m^2}{2}\varphi^2 \right),
\end{eqnarray}
 we have the
following set of equations instead of (\ref{aa}) 
\begin{eqnarray}
\label{aa1}
(\Box+m^2) \varphi &=&  0, \nonumber \\
{\cal R}_{\mu\nu} &=& 2D_\mu S_\nu - \frac{4}{3}g_{\mu\nu} S_\rho S^\rho 
+\frac{1}{2} \partial_\mu\varphi \partial_\nu\varphi 
-\frac{1}{2} g_{\mu\nu} m^2\varphi^2, \\
D_\mu S^\mu &=& \frac{3}{4}m^2\varphi^2, \nonumber \\
\tilde{K}_{\alpha\beta\gamma} &=& 0. \nonumber
\end{eqnarray}
The equation for the trace of the torsion tensor is different than the one of
the vacuum case, we have that massive scalar field 
couples to torsion in a different way than the massless one.
In contrast to the massless case, the equations (\ref{aa1}) do not admit as
solution $S_\mu=0$ for non-vanishing $\varphi$ (Again for particle-like
solutions we have $\phi=0$ and $S_\mu=0$).
This is in disagreement with the traditional belief that torsion must be
governed by spin distributions. We will return to this point in the last 
section.

\subsection{Gauge fields}

We need to be careful with the use of MCP to gauge fields. We will restrict
ourselves to the abelian case in this work,
 non-abelian gauge fields will bring some
technical difficulties that  will not contribute to the understanding
of the basic problems  of gauge fields on Riemann-Cartan spacetimes. 

Maxwell field can be described by the differential 
$2$-form
\begin{equation}
F = dA = d(A_\alpha dx^\alpha) = \frac{1}{2}F_{\alpha\beta}dx^\alpha
\label{form}
\wedge dx^\beta,
\end{equation}
where $A$ is the (local) potential $1$-form, and 
$F_{\alpha\beta}=\partial_\alpha A_\beta- \partial_\beta A_\alpha$ is the
usual electromagnetic tensor. It is important to stress that the
forms $F$ and 
$A$ are covariant objects in any differentiable manifolds. Maxwell equations
can be written in Minkowski spacetime in terms of exterior calculus as
\begin{eqnarray}
\label{maxeq}
dF&=&0, \\
d {}^*\!F &=& 4\pi {}^*\! J, \nonumber
\end{eqnarray}
where ${}^*$ stands for the Hodge star operator and $J$ is the current
$1$-form, $J=J_\alpha dx^\alpha$. The first equation in (\ref{maxeq}) is
a consequence of the definition (\ref{form}) and of Poincar\'e's lemma. 
In terms of components, one has the familiar homogeneous and non-homogeneous
Maxwell's equations,
\begin{eqnarray}
\label{maxeq1}
\partial_{[\gamma} F_{\alpha\beta]} &=& 0, \\
\partial_\mu F^{\nu\mu} &=& 4\pi J^\nu, \nonumber
\end{eqnarray}
where ${}_{[\ \ \ ]}$ means antisymmetrization. We know also that the
non-ho\-mo\-ge\-nous equation follows from the extremals
of the following action
\begin{equation}
S = -\int \left(4\pi{}^*\!J\wedge A +\frac{1}{2} F \wedge {}^*\!F\right) =
    \int d^4x\left(4\pi J^\alpha A_\alpha - \frac{1}{4}
F_{\alpha\beta}F^{\alpha\beta} \right).
\label{actmink}
\end{equation}

If one tries to cast (\ref{actmink}) in a covariant way by using MCP in the
tensorial quantities, we have that Maxwell tensor will be given by
\begin{equation}
\label{tilda}
F_{\alpha\beta}\rightarrow
\tilde{F}_{\alpha\beta} = 
F_{\alpha\beta} - 2 S_{\alpha\beta}^{\ \ \rho}A_\rho,
\end{equation}
which explicitly breaks gauge invariance. With this analysis, one usually
arises the conclusion that gauge fields cannot interact minimally with
Einstein-Cartan gravity. We would stress another undesired
consequence, also related to the breaking of gauge symmetry, of the use of MCP
in the tensorial quantities. The homogeneous Maxwell equation, the
first of (\ref{maxeq1}), does not come from a Lagrangian, and of course,
if we choose to use 
MCP in the tensorial quantities we need also apply MCP to it. We get
\begin{equation}
\partial_{[\alpha} \tilde{F}_{\beta\gamma]} + 
2 S_{[\alpha\beta}^{\ \ \rho} \tilde{F}_{\gamma]\rho} = 0 ,
\label{falac}
\end{equation}
where $\tilde{F}_{\alpha\beta}$ is given by (\ref{tilda}). One can see that
(\ref{falac}) has no general solution for arbitrary 
$S_{\alpha\beta}^{\ \ \rho}$. Besides the breaking of gauge symmetry,
the use of MCP in the tensorial quantities also leads to a non consistent
homogeneous equation. 

However, MCP can be successfully applied for general gauge fields 
(abelian or not) in the differential form quantities \cite{saa2}.  As 
consequence, one has that the homogeneous equation is already in a
covariant form in any differentiable manifold, and that the covariant
non-homogeneous equations can be gotten from a Lagrangian obtained only by
changing the metric tensor and by
introducing the parallel volume element in the Minkowskian action
(\ref{actmink}). Considering the case where $J^\mu=0$, we have the
following action to describe the interaction of Maxwell fields and 
Einstein-Cartan gravity
\begin{equation}
\label{actmax}
S = S_{\rm grav} + S_{\rm Maxw} = -\int   \,d^4x e^{2\Theta} \sqrt{-g}
\left( 
R + \frac{1}{4}F_{\mu\nu}F^{\mu\nu}
\right).
\end{equation}
As in the previous cases, the equation $\tilde{K}_{\alpha\beta\gamma}=0$ 
follows from the extremals of (\ref{actmax}). 
The other equations will be
\begin{eqnarray}
\label{ee1}
&&\frac{e^{-2\Theta}}{\sqrt{-g}}\partial_\mu e^{2\Theta}\sqrt{-g} F^{\nu\mu}
=0, \nonumber \\
&& {\cal R}_{\mu\nu} = 2D_\mu S_\nu - \frac{4}{3}g_{\mu\nu}S_\rho S^\rho 
-\frac{1}{2} \left(F_{\mu\alpha}F^{\ \alpha}_\nu 
+\frac{1}{2}g_{\mu\nu} F_{\omega\rho}F^{\omega\rho} \right), \\
&& D_\mu S^\mu = -\frac{3}{8}F_{\mu\nu}F^{\mu\nu}. \nonumber 
\end{eqnarray}

One can see that the equations (\ref{ee1}) are invariant under the usual
$U(1)$ gauge transformations. It is also clear
from the equations (\ref{ee1}) that Maxwell fields can interact with the 
non-Riemannian structure of spacetime. Also, as in the massive
scalar case, the equations do not admit as solution $S_\mu=0$ for arbitrary
$F_{\alpha\beta}$, Maxwell fields are also sources to the spacetime torsion.
Similar results can be obtained also for non-abelian gauge fields\cite{saa2}.

\subsection{Fermion fields}

The Lagrangian for a (Dirac) 
fermion field with mass $m$ in the Minkowski spacetime
is given by
\begin{equation}
{\cal L}_{\rm F}=\frac{i}{2}\left(\overline{\psi}\gamma^a\partial_a\psi
- \left(\partial_a\overline{\psi} \right)\gamma^a\psi \right)
- m\overline{\psi}\psi,
\label{fermion}
\end{equation}
where $\gamma^a$ are the Dirac matrices and 
$\overline{\psi}=\psi^\dagger\gamma^0$. Greek indices denote spacetime
coordinates (holonomic), and roman ones locally flat coordinates 
(non-holonomic). It is well known\cite{hehl}
that in order to cast (\ref{fermion}) in a covariant way, one needs to
introduce the vierbein field, $e^\mu_a(x)$, and
to generalize the Dirac matrices,
$\gamma^\mu(x) = e^\mu_a(x)\gamma^a$. The partial derivatives also must be 
generalized with the introduction of the spinorial connection $\omega_\mu$,
\begin{eqnarray}
\partial_\mu\psi \rightarrow 
 \nabla_\mu\psi &=& \partial_\mu\psi+ \omega_\mu \psi, \nonumber \\
\partial_\mu\overline{\psi} \rightarrow 
 \nabla_\mu\overline{\psi} &=& \partial_\mu\overline{\psi} - 
\overline{\psi}\omega_\mu,
\end{eqnarray}
where the spinorial connection is given by
\begin{eqnarray}
\label{spincon}
\omega_\mu &=& \frac{1}{8}[\gamma^a,\gamma^b]e^\nu_a\left(
\partial_\mu e_{\nu b} -\Gamma^\rho_{\mu\nu}e_{\rho b}\right) \\
&=& \frac{1}{8}\left(
\gamma^\nu\partial_\mu\gamma_\nu - \left(\partial_\mu\gamma_\nu \right)
\gamma^\nu - \left[\gamma^\nu,\gamma_\rho \right] \Gamma^\rho_{\mu\nu}
\right). \nonumber
\end{eqnarray}
 The last
step, according to our hypothesis, shall be 
the introduction of the parallel
 volume element, and after that one
gets the following action for fermion fields on RC manifolds
\begin{equation}
S_{\rm F} = \int d^4x e^{2\Theta}\sqrt{-g}\left\{ 
\frac{i}{2}\left(\overline{\psi}\gamma^\mu(x)\nabla_\mu\psi -
\left(\nabla_\mu\overline{\psi}\right)\gamma^\mu(x)\psi \right)
-m\overline{\psi}\psi
\right\}.
\label{fermioncov}
\end{equation}

Varying the action (\ref{fermioncov}) with respect to $\overline{\psi}$ one
obtains:
\begin{equation}
\frac{e^{-2\Theta}}{\sqrt{-g}}\frac{\delta S_{\rm F}}{\delta\overline{\psi}} =
\frac{i}{2}\left(\gamma^\mu\nabla_\mu\psi + \omega_\mu\gamma^\mu\psi \right)
-m \psi + \frac{i}{2}\frac{e^{-2\Theta}}{\sqrt{-g}} \partial_\mu
e^{2\Theta}\sqrt{-g}\gamma^\mu\psi = 0.
\end{equation}
Using the result
\begin{equation}
[\omega_\mu,\gamma^\mu]\psi = - \left( 
\frac{e^{-2\Theta}}{\sqrt{-g}}\partial_\mu e^{2\Theta}\sqrt{-g}\gamma^\mu
\right)\psi,
\end{equation}
that can be check using  (\ref{spincon}), 
(\ref{gammacontr}),  and 
properties of ordinary Dirac matrices and of the vierbein field,
 we get the following equation for $\psi$ on a RC spacetime:
\begin{equation}
\label{psi}
i\gamma^\mu(x)\nabla_\mu\psi - m\psi =0.
\end{equation}
The equation for $\overline{\psi}$ can be obtained in a similar way,
\begin{equation}
\label{psibar}
i \left( \nabla_\mu\overline{\psi}\right) \gamma^\mu(x)
 + m\overline{\psi} = 0.
\end{equation}

We can see that the equations (\ref{psi}) and (\ref{psibar}) are the same
ones that arise from MCP used in the minkowskian equations of motion. In the
usual EC theory, the equations obtained from the action principle do not
coincide with the equations gotten by generalizing the minkowskian 
ones. This is another new feature of the proposed model.

The Lagrangian that describes the interaction of fermion fields with the
Einstein-Cartan gravity is
\begin{eqnarray}
\label{actferm}
S &=& S_{\rm grav} +
S_{\rm F} \\ &=& - \int d^4x e^{2\Theta}\sqrt{-g} \left\{
R - \frac{i}{2}\left(\overline{\psi}\gamma^\mu\partial_\mu\psi -
\left(\partial_\mu\overline{\psi}\right)\gamma^\mu\psi 
\right.\right. \nonumber \\
&& \ \ \ \ \ \ \ \ \ \ \  
+ \left.\left.
\overline{\psi}\left[\gamma^\mu,\omega_\mu\right] \psi\right)
  + m\overline{\psi}\psi \right\} \nonumber \\
&=& - \int d^4x e^{2\Theta}\sqrt{-g} \left\{
R - \frac{i}{2}\left(\overline{\psi}\gamma^\mu\partial_\mu\psi -
\left(\partial_\mu\overline{\psi}\right)\gamma^\mu\psi 
\right.\right. \nonumber \\
&& \ \ \ \ \ \ \ \ \ \ \ 
+ \left.\left. 
\overline{\psi}\left[\gamma^\mu,\tilde{\omega}_\mu\right] \psi\right)  
 -\frac{i}{8}\overline{\psi}\tilde{K}_{\mu\nu\omega}
\gamma^{[\mu}\gamma^\nu\gamma^{\omega]} \psi 
  + m\overline{\psi}\psi \right\},\nonumber
\end{eqnarray}
where it was used that 
$\gamma^a\left[\gamma^b,\gamma^c\right]+
\left[\gamma^b,\gamma^c\right]\gamma^a=
2\gamma^{[a}\gamma^b\gamma^{c]}$, and that
\begin{equation}
\omega_\mu = \tilde{\omega}_\mu +\frac{1}{8}K_{\mu\nu\rho}
\left[\gamma^\nu,\gamma^\rho\right],
\end{equation}
where $\tilde{\omega}_\mu$ is the 
Riemannian spinorial connection, calculated by
using the Christoffel symbols instead of the full connection in 
(\ref{spincon}).

The peculiarity of fermion fields is that one has a non-trivial equation
for $\tilde{K}$ from (\ref{actferm}). 
The Euler-Lagrange equations for $\tilde{K}$ is given by
\begin{eqnarray}
\frac{e^{-2\Theta}}{\sqrt{-g}} \frac{\delta S}{\delta\tilde{K}}  = 
\tilde{K}^{\mu\nu\omega} + \frac{i}{8}\overline{\psi}
\gamma^{[\mu}\gamma^\nu\gamma^{\omega]}\psi = 0.
\label{ka}
\end{eqnarray}
Differently from the previous cases, we have that the  traceless part of
the contorsion tensor,
$\tilde{K}_{\alpha\beta\gamma}$, is proportional to the spin
distribution. It is still zero outside
matter distribution, since its equation is an algebraic one, it does not
allow propagation. The other equations follow from the extremals of
(\ref{actferm}). The main difference between these equations and the usual
ones obtained from standard EC gravity, is that in the present case one 
has non-trivial solution for the trace of the torsion tensor, that is
derived from $\Theta$. In the standard EC gravity, the torsion tensor is
a totally anti-symmetrical tensor and thus it has a vanishing trace.

\section{Final remarks}

In this section, we are going to discuss the role of the 
 condition (\ref{pot}) and the source for torsion in the proposed model.
The condition  (\ref{pot}) is the necessary
condition in order to be possible the definition of a parallel
volume element on a manifold. Therefore, we have that our approach is 
restrict to spacetimes which admits such volume elements. 
We automatic have this restriction if we wish to use
MAP in the sense discussed in Section II.
Although it is not clear how to get EC gravity equations without
using a minimal action principle, we can speculate about matter fields
on spacetimes not obeying (\ref{pot}). Since it is not equivalent
to use MCP in the equations of motion or in the action formulation, we
can forget the last and to cast the equations of motion for matter
fields in a covariant way directly. It can be done easily, as example,
for scalar fields\cite{saa1}. We get the equation (\ref{e3}),
which is, apparently, a consistent equation. However, we need to define
a inner product for the space of the solutions of (\ref{e3})
\cite{dewitt}, and we are able to do it only if (\ref{pot}) holds.
We have that the dynamics of matter fields requires some restrictions 
to the non-riemannian structure of spacetime, namely, the condition
(\ref{pot}). This is more evident for gauge fields, where 
(\ref{pot}) arises directly as an integrability condition for the
equations of motion \cite{saa2}. It seems that condition (\ref{pot}) cannot
be avoided.

We could realize from the matter fields studied that the trace of the
torsion tensor is not directly related to spin distributions. This is a
new feature of the proposed model, and we naturally arrive to the 
following question: What is the source of torsion? The situation for the
traceless part of the torsion tensor is the same that one has in the
standard EC theory, only fermion fields can be sources to it. As to the
trace part, it is quite different.
 Take for example $\tilde{K}_{\alpha\beta\gamma}=0$, that corresponds to
scalar and gauge fields. 
In this case, using the definition of the energy-momentum tensor
\begin{equation}
\frac{e^{-2\Theta}}{\sqrt{-g}}
\frac{\delta S_{\rm mat}}{\delta g^{\mu\nu}} = -\frac{1}{2}T_{\mu\nu},
\end{equation}
and that for scalar and gauge fields we have
\begin{equation}
\frac{e^{-2\Theta}}{\sqrt{-g}}
\frac{\delta S_{\rm mat}}{\delta \Theta} = 2 {\cal L}_{\rm mat},
\end{equation}
one gets
\begin{equation}
D_\mu S^\mu = \frac{3}{2}
\left(  {\cal L}_{\rm mat} - \frac{1}{2}T
\right),
\end{equation}
where $T$ is the trace of the energy-momentum tensor.
The quantity between parenthesis, in general, has nothing to do with spin, and 
it is the source for a part of the torsion, confirming that in
our model part of torsion is not determined by spin distributions. See
also \cite{H1} for a discussion on possible source terms to the torsion.

This work was supported by FAPESP. The author wishes to thank an anonymous
referee for pointing out the reference \cite{H1}.

\end{document}